\documentclass[%
reprint,
 amsmath,amssymb,
 aps,
 superscriptaddress,
floatfix,
]{revtex4-2}

\usepackage{graphicx}
\usepackage{dcolumn}
\usepackage{bm}

\usepackage{xcolor}

\usepackage {romannum}

\usepackage {hyperref}
\usepackage{float}
\usepackage{lineno}
\usepackage{booktabs}

\begin{document}






\title{two-dimensional half Chern-Weyl semimetal with multiple screw axes}%

\author{Wei Xu}%
\thanks{These two authors contributed equally to this work.}
\affiliation{Department of Physics, East China University of Science and 
Technology, Shanghai 200237, China}
\author{Jiawei Yi}%
\thanks{These two authors contributed equally to this work.}
\affiliation{Department of Physics, East China University of Science and 
Technology, Shanghai 200237, China}
\author{Hao Huan}%
\affiliation{State Key Laboratory of Surface Physics and Key Laboratory of Computational 
Physical Sciences (MOE) $\&$ Department of Physics, Fudan University, Shanghai 200433, China}
\author{Bao Zhao}%
\affiliation{State Key Laboratory of Surface Physics and Key Laboratory of Computational 
Physical Sciences (MOE) $\&$ Department of Physics, Fudan University, Shanghai 200433, China}
\affiliation{Shandong Key Laboratory of Optical Communication Science and Technology, 
School of Physics Science and Information Technology, Liaocheng University, Liaocheng 252059, China}  
\author{Yang Xue}%
\email[E-mail: ]{xuey@ecust.edu.cn}
\affiliation{Department of Physics, East China University of Science and 
Technology, Shanghai 200237, China}
\author{Zhongqin Yang}%
\affiliation{State Key Laboratory of Surface Physics and Key Laboratory of Computational 
Physical Sciences (MOE) $\&$ Department of Physics, Fudan University, Shanghai 200433, China}

\date{\today}%

\begin{abstract}
Half topological states of matter and two-dimensional (2D) magnetism have gained much attention recently. In this paper, we propose a special topological semimetal phase called a 2D half Chern-Weyl semimetal (HCWS), which is a 2D magnetic Weyl semimetal bound to the half Chern insulator phase by symmetry, and the two phases can be converted to each other by manipulating the magnetization direction. We provide the symmetry conditions to realize this state in 2D systems with multiple screw axes. Tight-binding models with multiple basis and a predicted 2D material, monolayer TiTe, are shown as the concrete examples for HCWSs. The TiTe monolayer was shown to have a high ferremagnetic Curie temperature (\textasciitilde966 K) as well as a Coulomb correlation-enhanced spin-orbit coupling (SOC), and further demonstrates the effect of correlation-enhanced SOC on magnetocrystalline anisotropy energy and energy gap opening. Our work reveals a state with switchable and spin-resolved half body charge currents as well as half boundary charge currents, and will provide a platform for novel and high-performance topological spintronics devices.  
\end{abstract}

\maketitle

\textit{Introduction.}---Two-dimensional (2D) half topological states, possessing fully-spinpolarized topologically nontrivial energy bands near Fermi level ($\text{E}_\text{F}$), have gained much attention \cite{SpinPolarized_Nodalloop2019,2D_WeylHalf_shengyuan_yang,WeylMonoloop_Half,2D_WeylHalf_NiCS3,NPG_halfChernInsulators2018,Half_QAH_Valley_Huanhao,FeI_MAE_strain,xuyong_QAH_LiFeSe_2020,Ni2I2_Liulei}. Two outstanding representatives, 2D half Weyl semimetals \cite{2D_WeylHalf_shengyuan_yang,2D_WeylHalf_NiCS3,SpinPolarized_Nodalloop2019,WeylMonoloop_Half} and half Chern insulators (HCIs) \cite{NPG_halfChernInsulators2018,Half_QAH_Valley_Huanhao,FeI_MAE_strain,xuyong_QAH_LiFeSe_2020,Ni2I2_Liulei}, with half-metallic linear bulk bands and half-metallic chiral edge states respectively, show potential applications in high-performance and thin spintronics devices. Unlike in the 3D case, Weyl points in 2D systems must be protected by additional crystal symmetries due to the loss of topological protection by dimension reduction \cite{2D_WeylHalf_shengyuan_yang}. In addition, for ferromagnetic (FM) materials, the crystal symmetry is related to the magnetization direction. Thus, the half Weyl points in 2D materials are governed by their magnetization direction, which gives such systems more diverse manipulation means based on regulating the magnetization direction.

Besides the crystal symmetry, spin-orbit coupling (SOC) and electron correlation are also at the core of 2D half topological states. Specifically, the long-range FM order in 2D half topological states is stabilized via SOC-induced magnetocrystalline anisotropy, lifting the Mermin-Wagner restriction \cite{Mermin_Wagner_theorem}. Electron correlation is also a significant effect in 2D magnetic systems which may enhance their SOC \cite{U_enhanced_SOC_MAE}. Thus, energy gap opened by SOC and magnetocrystalline anisotropy energy (MAE) are closely related to electron correlation. 2D half Weyl semimetals provide an excellent platform for studying the entangled physics of topological properties, crystal symmetry, SOC, and electron correlation.   

Here we draw attention to a peculiar class of Weyl semimetals called 2D half Chern-Weyl semimetals (HCWSs), which are 2D magnetic Weyl semimetals capable of opening a nontrivial global energy gap characterized by Chern numbers by changing their magnetization direction. As shown in Fig. 1, there are two different transport channels in HCWSs that can both transport fully spin-polarized charge currents. The two channels can be switched by manipulating the magnetization direction in HCWSs and the direction of spin in these two channels are different, parallel to the different magnetization directions. These novel properties show their great application in spintronics.

\begin{figure}[htbp]
  \centering
\includegraphics[]{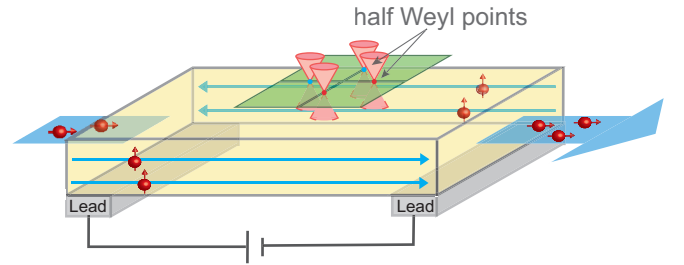}
\caption{Schematic diagram of the two types of transport channels of HCWSs. Both channels can transport fully spin-polarized charge currents, where the thick blue arrow indicates bulk transport channel and the thin blue arrows represent edge transport channels. The spin orientations in them are different, denoted by red arrows. The green plane is the BZ, and the red and blue points on it are chiralities of half Weyl points, red for positive and blue for negative.
}
\label{1}
\end{figure}

By generalizing nonmagnetic Dirac semimetals with multiple screw axes \cite{Cat_cradle_Dirac,abc,Weyl_Dirac_Screw} to magnetic systems, two types of hourglasslike, as well as nodal line 2D HCWSs are obtained. Futhermore, we predict a 2D magnets with high ferromagnetic Curie temperature of 966 K, monolayer TiTe, as a concrete candidates for realizing the 2D hourglasslike HCWSs. The two pairs of half Weyl points in TiTe are protected by a screw symmetry and a glide symmetry, respectively. The systems transform into HCIs after breaking the pristine nonsymmorphic symmetries by rotating the magnetization direction. The Coulomb correlation-enhanced and magnetization direction-dependent SOC, MAE, and energy gap opening are explicitly demonstrated in TiTe. 
Our work not only reveals a state of matter and its interesting physical connotations but also offers promising material platforms for novel topological spintronics applications. 

\textit{2D HCWSs and HCIs from band inversion}---We first analyze the conditions needed for realizing 2D half Weyl semimetals in the systems with multiple screw axes. Then give the requirements for non-trivial gap opening in these 2D half Weyl semimetals. For all 80 nonmagnetic layer groups, there are total 11 layer groups with two screw axes along the axial direction \cite{Cat_cradle_Dirac}. Later, we will focus on the layer groups with inversion symmetry among the above 11 layer groups and take the representative case $pmmn$ (No. 46) as our starting point. The generators of layer group $pmmn$ contain two screw operations ($\widetilde{C}_{2x}=\{C_{2x}\mid a/2\}$, $\widetilde{C}_{2y}=\{C_{2y}\mid b/2\}$), and an inversion $P$. By including collinear FM and SOC, the spatial rotation and spin rotation are locked globally, and the existence of the nonsymmorphic symmetries above will depend on magnetization direction. In the following, we discuss two typical FM cases with in-plane (x-axial) and out-of-plane (z-axial) magnetization vectors ($\bm{\hat{m}}$), respectively.

\textit{Case 1}---If $\bm{\hat{m}}$ is along x direction, the time-reversal symmetry $\mathcal{T}$ and the glide mirror symmetry $\widetilde{M}_{z}=P\widetilde{C}_{2x}\widetilde{C}_{2y}$ are broken, while the joint symmetry of these two symmetries is preserved,  denoted as $\Theta=\mathcal{T} \widetilde{M}_{z}$. $\Theta$ leaves the X point ($k_{x}=\pi, k_{y}=0$) and Y point ($k_{x}=0, k_{y}=\pi$) of Brillouin zone (BZ) invariant. On the other hand, we have $\Theta^{2}=e^{-i(k_{x}+k_{y})}$. Thus, at X and Y, $\Theta^{2}=-1$ will result in Krammers-like degeneracy. 

Due to the k-lines $\Gamma-X$ and $\Gamma-Y$ are invariant under $\widetilde{C}_{2x}$ and $\widetilde{M}_{x}=P\widetilde{C}_{2x}$, respectively. The eigenstates along $\Gamma-X$ and $\Gamma-Y$ can be labelled by the eigenvalues of $\widetilde{C}_{2x}$ ($\pm i e^{-i k_{x}/2}$) and $\widetilde{M}_{x}$ ($\pm i$), respectively. By considering the following relations
\begin{equation}
\begin{aligned}
  & \Theta \widetilde{C}_{2x}=-e^{i k_{y}}\widetilde{C}_{2x}\Theta \\
  & \Theta \widetilde{M}_{x} =-e^{i k_{x}}\widetilde{M}_{x}\Theta,
  \end{aligned}
\end{equation}
we can draw a conclusion that the two bands evolved from a Krammers-like degenerate point at X (Y) will carry the same eigenvalues of $\widetilde{C}_{2x}$ ($\widetilde{M}_{x}$) along $\Gamma-X$ ($\Gamma-Y$) (details see Supplemental Material \cite{SMs}). As shown in Figs. 2(c-e), if band inversion happens at the $\Gamma$ point between two pairs of bands, two types of hourglasslike Weyl semimetals emerge, the hourglasslike Weyl semimetals with a pair of Weyl points (Figs. 2(d) and 2(e)) and with two pairs of Weyl points (Fig. 2(c)). Noticing that the $\widetilde{C}_{2x}$ and $\widetilde{M}_{x}$ do not flip the x-direction spins, thus the above obtained Weyl fermions are all in single spin channel. 

\begin{figure}[htbp]
  \centering
\includegraphics[]{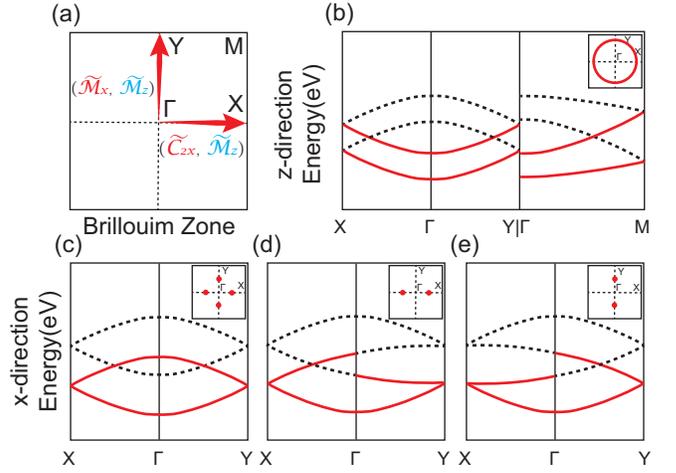}
\caption{(a) The first BZ with high-symmetry points and lines. The $\Gamma$-X (Y) lines are invariant under $\widetilde{C}_{2x}$ ($\widetilde{M}_{x}$) or $\widetilde{M}_{z}$ ($\widetilde{M}_{z}$) when magnetization direction is parallel to x or z, respectively. (b)-(e) The Schematic band structures for four distinctive half Weyl semimetal states of $pmmn$ layer group with FM order along x and z directions, respectively. Red solid lines and black dot lines denote the different eigenvalues of the symmetry operations on the high-symmetry lines given in (a). Each inset shows the band-crossing points inside the BZ.
}
\label{2}
\end{figure}

\textit{Case 2}---If $\bm{\hat{m}}$ is in the z direction, a similar discussion can be carried out using the preserved symmetries $\widetilde{M}_{z}$, $\Theta_{x}=\mathcal{T} \widetilde{C}_{2x}$, and $\Theta_{y}=\mathcal{T} \widetilde{C}_{2y}$. $\Theta_{x}$ and $\Theta_{y}$ leave the boundaries of BZ $X-M$ and $Y-M$ invariant and result in a Krammers-like degeneracy along them, respectively. Meanwhile, eigenstates inside the whole BZ can be denoted by eigenvalues of $\widetilde{M}_{z}$, which leaves the BZ invariant. By considering the following relations
\begin{equation}
\begin{aligned}
  & \widetilde{M}_{z}\Theta_{x} =-e^{i k_{y}}\Theta_{x} \widetilde{M}_{z}\\
  & \widetilde{M}_{z}\Theta_{y} =-e^{i k_{x}}\Theta_{y} \widetilde{M}_{z},
  \end{aligned}
\end{equation}
we make a conclusion that the two bands, evolved from a Krammers-like degenerate point at boundaries of the BZ, will carry the different eigenvalues of $\widetilde{M}_{z}$ inside the whole BZ. Combining band inversion at the $\Gamma$ point, this conclusion will lead to a nodal line Weyl semimetal, shown in Fig. 2(b).

Obviously, the Weyl points come in pairs for all the cases above, satisfying the no-go theorem \cite{no_go_theorem}.  The gap opening at a 2D Weyl point would give a topological charge of $\pm 1/2$ \cite{half_topo_charge_Weyl_point,wanxiangang_Weyl_SemimetalFermiarc2011}. Moreover, the topological charges of two Weyl points connected by inversion $P$ must be identical and will contribute to $\pm 1$ Chern numbers for the case in Figs. 2(d) and 2(e) (Berry curvature $\Omega$ is even under $P$). To obtained non-zero Chern numbers for the cases with multiple pairs of Weyl points (Figs. 2(b) and 2(c)), additional symmetry need to be included, such as $\widetilde{C}_{4z}$, to ensure that the topological charges of different Weyl points do not completely cancel out.

To show some examples, we built square lattice symmetry-enforced tight-binding (TB) models \cite{MagneticTB} for different symmetry-allowed d-orbital doublets with the same spin. By considering in-plane and out-of-plane magnetization, respectively, and by tuning the parameters of TB models, all possible half Weyl semimetals under different basis are listed in Table \ref{tab:TB}. The details of TB models are given in Supplemental Material \cite{SMs}.

\begin{table}[htbp]
  \caption{\label{tab:TB}
  The possible half Weyl semimetal states obtained from symmetry-enforced TB models on a square lattice with different FM orientations. The basis sets are listed in first column.}
  \begin{ruledtabular}
  \begin{tabular}{ccc}
 &$FM_{x}$\footnotemark[1]&$FM_{z}$\\
  \hline
($d_{xy}$, $d_{z^{2}}$) & hourglasslike-I\footnotemark[2] & nodal line \\
($d_{x^{2}-y^{2}}$, $d_{z^{2}}$) & hourglasslike-I & nodal line \\
($d_{x^{2}-y^{2}}$, $d_{xy}$) & hourglasslike-I or II & nodal line \\
($d_{xz}$, $d_{yz}$) & hourglasslike-I or II & \\
  \end{tabular}
  \end{ruledtabular}
  \footnotetext[1]{The subscripts x and z refer to the magnetization directions.}
  \footnotetext[2]{Roman numerals refer to the number of Weyl point pairs in an hourglasslike Weyl semimetal, I for one pair and II for two pairs.}
  \end{table}

\textit{material examples}---Next, we reveal a concrete material example of 2D HCWSs, monolayer TiTe, whose geometric structure is displayed in Fig. 3(a). Its unit cell has $p4/nmm$ space group (No. 129) symmetry and comprises two Ti and two Te atoms. The crystal structure has two screw axes: $\widetilde{C}_{2x}=\{C_{2x}\mid (a/2,0,0)\}$ and $\widetilde{C}_{2y}=\{C_{2y}\mid (0,a/2,0)\}$. The inversion center is located at the crossing point of the two screw axes. The fully optimized lattice parameters are $a=b=4.33$ \AA for U=3 eV  (Fig. 3(b)). The phonon spectra (Fig. S1(a)) without imaginary frequency mode manifests the dynamical stability of the monolayer TiTe. The thermal stability of monolayer TiTe is further confirmed by first-principles molecular-dynamics (MD) simulation (Fig. S1(b)), which shows that the structure remains intact at a temperature of 300 K after 10 ps. 

\begin{figure}[htbp]
  \centering
\includegraphics[trim = 0.0cm 0cm 0cm 0cm,clip]{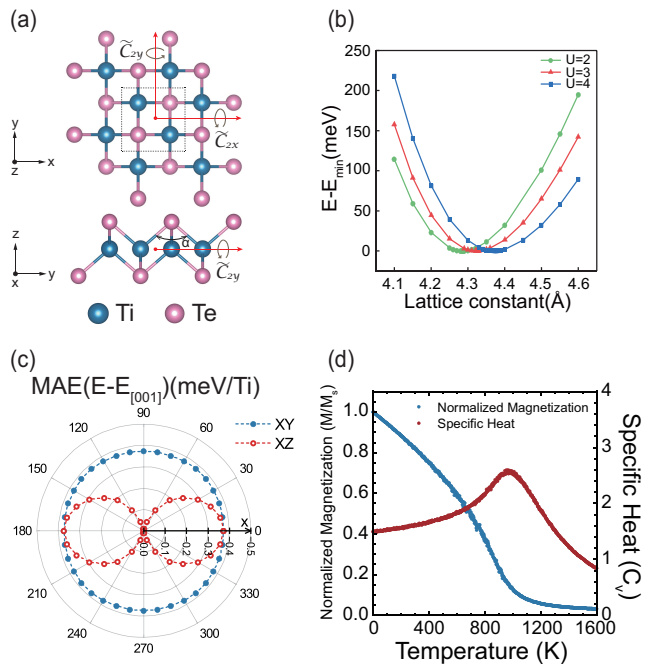}
\caption{(a) Top and side view of 2D TiTe monolayer, the blue and pink balls represent Ti and Te atoms, respectively. Dashed lines show unit cell. The red arrows indicate the screw axes of $\widetilde{C}_{2x}$ and $\widetilde{C}_{2y}$. The red dot represents the inversion center. (b) The total energies as a function of the lattice constant for TiTe monolayer with U=2|3|4 eV, respectively, in which the minimum total energy at the equilibrium lattice constant for each case is set as energy zero. (c) The MAE by rotating the spin within x-z and x-y planes, respectively. (d) The normalized average magnetic moment (blue curve) and specific heat (red curve) versus temperature for the TiTe monolayer. The results in (c) and (d) are calculated with U=3 eV.}
\label{3}
\end{figure}

\begin{table}[htbp]
  \caption{\label{tab:table1}The relative total energy (meV per unit cell) of four magnetic configurations shown in Fig. S2, magnetic coupling parameters of $ J_1 $ and $J_2$ (meV), anisotropy parameter D (meV per Ti atom), and critical temperature $T_c$ (K) obtained from the Heisenberg model for TiTe monolayer with U=3 and 5 (eV), respectively.}
  \begin{ruledtabular}
  \begin{tabular}{ccccccccc}
    {} & FM & AFM-N & AFM-L & AFM-Z & $J_1$ & $J_2$ & D & $T_c$ \\
    \hline
    \specialrule{0em}{2pt}{2pt}
    U=3 & 0 & 557.1 & 586.4 & 332.5 & 69.64 & 38.48 & -0.37 & 966 \\
    U=5 & 0 & 628.8 & 689.1 & 25.1 & 78.60 & 46.83 & 5.70 & 1148 \\
  \end{tabular}
  \end{ruledtabular}
  \end{table}

To determine the magnetic ground state, the $2 \times 2 \times 1$ supercell of the monolayer TiTe with four magnetic configurations and different Hubbard U values (2-5 eV) are considered (see Fig. S2). It is found that the monolayer TiTe maintains a FM ground state for all considered U values (Table II). The magnetic moment of TiTe is 2$\mu_{B}$ per Ti atom from our DFT calculations. It can be well understood by analyzing the valence electron configuration of Ti atom ($3d^{2}4s^{2}$), which becomes $\text{Ti}^{2+}$ ($3d^{2}$ configuration) after transfer $2e^{-}$ to Te atom, confirmed by the Bader charge calculations \cite{Bader_charge}. Unpaired $2e^{-}$ electrons prefer to have high spin alignment with 2$\mu_{B}$ based on Hund's rules, which can be seen in the partial densities of states (DOSs) of TiTe (Fig. 4(c)). The microscopic mechanism of FM ground state for TiTe is explained by a Ti-Te-Ti FM super-exchange interaction according to the Goodenough-Kanamori-Anderson rules \cite{goodenough_kanamori_anderson,GKA_A,GKA_K,GKA_G}, with a bond angle of Ti-Te-Ti ($\alpha=98.63^{\circ}$ in Fig. 3(a)) close to the $90^{\circ}$.

\begin{figure}[htbp]
  \centering
\includegraphics[trim = 0.9cm 0.5cm 0cm 0cm,clip]{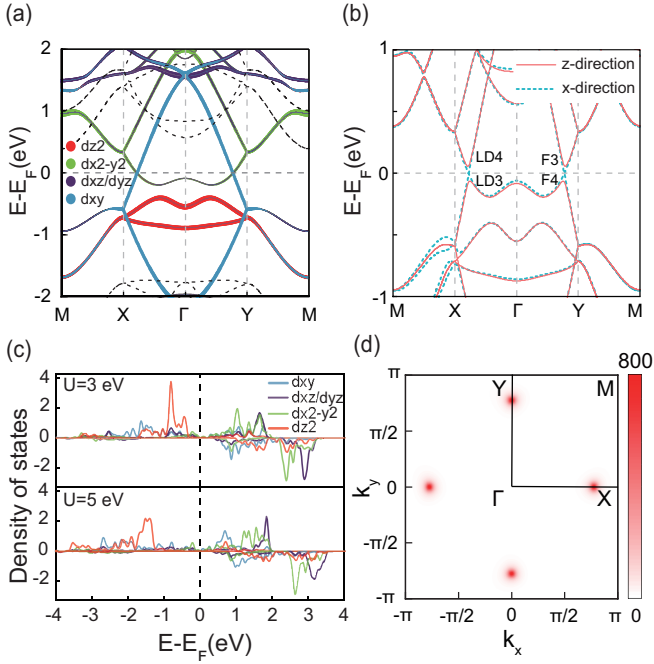}
\caption{(a) The orbital resolved band structures of TiTe monolayer without SOC. The orbital projections are only applied for the bands in spin-up channel and the spin-down bands are plotted with black dashed lines. (b) Band structures with SOC of TiTe monolayer with x-magnetization and z-magnetization, respectively. The irreducible representations of the two bands crossing the $\text{E}_\text{F}$ for x-magnetization case are displayed. (c) The projected densities of states for the TiTe monolayer with two U values, respectively, in which the $d_{xz}$ and $d_{yz}$ orbitals are degenerate. (d) Distribution of the Berry curvatures for the TiTe monolayer with the magnetization direction along the z axis.}
\label{4}
\end{figure}

The MAE derived from the SOC plays a crucial role  for a 2D material keeping in a long-range FM ground state \cite{Magnetism_Nanoscale2006}. The MAE of TiTe monolayer in the xz and xy planes as functions of the magnetization angle $\theta$ is depicted in Fig. 3(c). By comparing the energy difference of the two planes, we find that the magnetic easy axes of the TiTe monolayer prefer the in-plane direction with the $\text{U}=3.0$ eV. This trend of MAE remains consistent in the range of $\text{U} = 2-3.4 $ eV. With increasing U, the magnetic anisotropy direction changes from the in-plane to the out-of-plane, and the critical U value is about 3.4 eV (Fig. 5(c)). The mechanism for this interesting U-correlated MAE phenomenon will be explained latter. To evaluate the $T_C$ of the TiTe monolayer, a Heisenberg model is built as \cite{VP_weiren}

\begin{equation}
  H_{0}=-J_{1}\sum_{<i,j>}\boldsymbol{S_{i}}\boldsymbol{\cdot S_{j}}-J_{2}\sum_{<<i,j>>}\boldsymbol{S_{i}}\boldsymbol{\cdot S_{j}}-D\sum_{i}\mid S_{i}^{e}\mid^{2},
  \end{equation}
where $\boldsymbol{S_{i}}$ is the spin vector, $\boldsymbol{S_{i}^{e}}$ is the spin component along the easy axis, $J_{i}$ ($i=1,2$) and $D$ denote the strengths for exchange interaction and anisotropy, respectively. Both the first and second terms can be regarded as isotropic. Both $J_{i}$ and $D$ can be extracted from the first-principles calculations with the equations in Supplemental Material \cite{SMs} and are listed in Table II. As shown in Fig. 3(d), the Curie temperature for the FM state can be estimated as $T_C \thickapprox 966 \text{K}$, which is significantly higher than the room temperature and also than that of the 2D star magnetic material of the CrI3 monolayer (45 K \cite{CrI3_45K} or 95 K \cite{CrI3_95K}).

For now, the magnetic ground state for TiTe monolayer is confirmed as FM, satisfying the prerequisites for 2D HCWSs. By calculating the electronic band structures of TiTe monolayer without SOC, one observes two pairs of fully spin-polarized twofold band crossing points symmetrically distributed on the x-axis and the y-axis in Fig. 4(a), which are related with each other by $\widetilde{C}_{4z}=\{C_{4z}\mid (a/2,0,0)\}$ and represent 2D Weyl points. This half-semiconducting feature is robust to the change in Hubbard U (Fig. S3). 

Considering the SOC effect, the stability of the 2D Weyl points depends on the magnetization directions. For the x-axial magnetization, these half Weyl points can be protected by preserved nonsymmorphic symmetries $\widetilde{C}_{2x}$ and $\widetilde{M}_{x}$, shown in Fig. 4(b), consistent with the discussion in Fig. 2(c). The marked different irreducible representations of LD3 and LD4 (F3 and F4) confirm the robustness of the half Weyl points in TiTe with the in-plane-axial magnetization. As shown in Fig. 5(a), the edge state of TiTe starts from the Weyl point projection on $-X-\Gamma$ and ends at the projection of another Weyl point on $X-\Gamma$, very similar to the Fermi arc in 3D Weyl semimetals \cite{wanxiangang_Weyl_SemimetalFermiarc2011,exp_TaAs_Weylsemi,TaAs_WeylSemi_Weng,HgCr2Se4_WyelSemi}.

\begin{figure}[htbp]
  \centering
\includegraphics[trim = 6.9cm 0.9cm 0cm 0cm,clip]{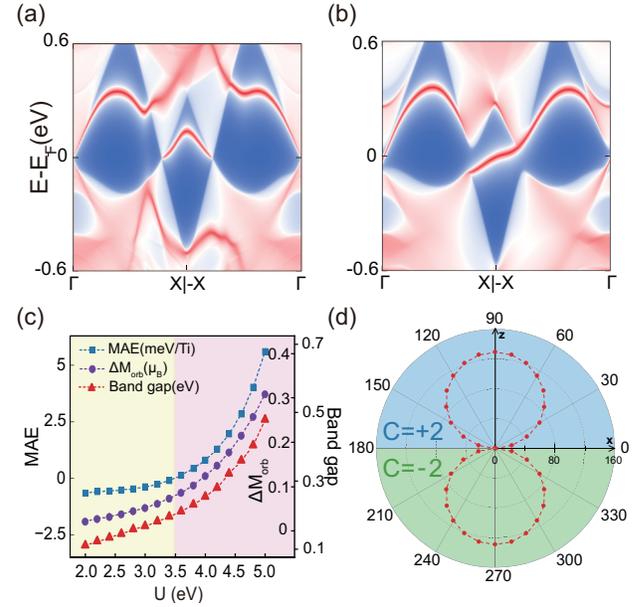}
\caption{(a) and (b)  The edge states of a semi-infinite TiTe monolayer cut along the [100] direction with x-magnetization and z-magnetization, respectively. (c) The MAE, orbital moment anisotropy ($\Delta M_{orb} =|M_{orb}^{[001]}|-|M_{orb}^{[100]}| $), and maximal band gap as functions of U. (d) The band gap of the TiTe monolayer with the magnetization direction varying from $\theta=0^\circ$ to $\theta={360}^\circ$ in the xz plane. The green and blue areas indicate the monolayer with Chern numbers of -2 and 2, respectively. 
}
\label{5}
\end{figure}

To characterize the low-energy band structures for the 2D HCWSs state, we construct a $k \cdot p$ effective model. With x-axis magnetization and SOC, the effective Hamiltonian for half Weyl points in $-X-X$ is subjected to the magnetic little co-group $m^{\prime}m2^{\prime}$ of $-X-X$, with two generators, $\widetilde{C}_{2x}$ and $\mathcal{T}\widetilde{C}_{2y}$. The symmetry constraints are given by

\begin{equation}
\begin{aligned}
  \widetilde{C}_{2x}^{-1}H\left(q_{x},q_{y}\right)\widetilde{C}_{2x} & =H\left(q_{x},-q_{y}\right)\\
  \left(\mathcal{T}\widetilde{C}_{2y}\right)^{-1}H\left(q_{x},q_{y}\right)\left(\mathcal{T}\widetilde{C}_{2y}\right) & =H^{\ast}\left(q_{x},-q_{y}\right),
  \end{aligned}
\end{equation}
where $\mathbf{q}$ is measured from the Weyl points in $-X-X$. In the basis of the irreducible representations LD3 and LD4, we find that to linear order in $\mathbf{q}$, the effective model takes the form of the 2D Weyl model,

\begin{equation}
H=A\left(\tau q_{x}\right)\sigma_{z}+B\left( q_{y}\right)\sigma_{x}+C\left( q_{x}\right)\sigma_{0},
\end{equation}
where $A\left(\tau q_{x}\right)=\tau(c_{1}q_{x}-c_{2}q_{x})+c_{3}-c_{4}$, $B\left(q_{y}\right)=c_{5}q_{y}$,  $C\left(q_{x}\right)=c_{1}q_{x}+c_{2}q_{x}+c_{3}+c_{4}$ ($c_{i=1-5}$ are constant parameters and $\tau=\pm$ for the Weyl points in $X-\Gamma$ or $-X-\Gamma$ respectively, $\sigma_{x}-\sigma_{z}$ are the Pauli matrixes, and $\sigma_{0}$ is the unit matrix. Thus, the low-energy electrons indeed resemble a pair of 2D half Weyl fermions located in $-X-X$ \cite{Magnetic_III}. 

Conversely, by rotating x-axis magnetization to z-axis magnetization, breaking the $\widetilde{C}_{2x}$ and $\widetilde{M}_{x}$ will remove the Weyl points and open an energy gap, as shown in Fig. 4(b). The gap opening at a 2D Weyl point would induce a finite Berry curvature $\Omega(\boldsymbol{q})=-2Im\langle \partial_{q_{x}} u_{v}\mid \partial_{q_{y}} u_{v} \rangle$, where $\mid  u_{v} \rangle$ is the eigenstate of the valence band. The integral of Berry curvature in a region around the Weyl point gives a topological charge of $\pm 1/2$ \cite{half_topo_charge_Weyl_point}. Since the four Weyl points are connected by preserved $\widetilde{C}_{4z}$, their topological charges have same sign, shown in Fig. 4(d), and contribute a non-zero Chern number ($C=2$), which suggests that two chiral edge states appear in the bulk gap (Fig. 5(b)). The edge states are confirmed to be fully-spinpolarized by calculations, showing great potential for dissipassionless spintronics.

The SOC effect of light $3d$ transition metal ions is usually small. While, a TiTe monolayer can open a band gap of up to 119 meV with $\text{U}=3$ eV, and the band gap increases significantly as the U value increases (\textasciitilde429 meV with $\text{U}=5$ eV), as shown in Fig. 5(c). The same U-enhancement effect also appears in the MAE of TiTe (Fig. 5(c)). These phenomena suggested a significant correlation-enhancing SOC effect \cite{U_enhanced_SOC_MAE} in TiTe. 

Correlation can enhance SOC by enlarging the effective SOC parameters for both in-plane orbital group $\{d_{xy},d_{x^{2}-y^{2}}\}$ and out-of-plane orbital group $\{d_{xz/yz},d_{z^{2}}\}$ in single spin channel \cite{SMs}.  For in-plane and out-of-plane groups, the enhanced effective SOC parameters $\lambda_{\parallel}$ and $\lambda_{\perp}$ are respectively given as

\begin{equation}
  \begin{aligned}
    \lambda_{\parallel}=\lambda+\frac{\left|\bar{L}_{z}\right|}{4\hbar^{3}}U_{eff}, \quad \lambda_{\perp}=\lambda+\frac{\left|\bar{L}_{x}\right|}{3\hbar^{3}}U_{eff},
    \end{aligned}
\end{equation}
where the $\lambda$ is pristine SOC parameter, $U_{eff}=U-J$ with $U$ and $J$ are the on-site Coulomb repulsion and exchange interaction parameters, and $\left|\bar{L}_{i}\right|=\sum_{m \in occ.}\left| \langle m \mid \hat{L}_{i} \mid m \rangle \right|, \textit{i=x or z}$ is the expectation of $\hat{L}_{i}$ over the occupied states. The $\left|\bar{L}_{x}\right|$ ($\left|\bar{L}_{z}\right|$) is also a function of $\lambda_{\perp}$ ($\lambda_{\parallel}$). Since the intrinsic SOC $\lambda$ is relatively small, $\left|\bar{L}_{x}\right|$ ($\left|\bar{L}_{z}\right|$) can be expressed as a function of $\lambda$: $\left|\bar{L}_{x}\right|=\kappa_{x}\lambda$ ($\left|\bar{L}_{z}\right|=\kappa_{z}\lambda$) and be solved self-consistently.

As shown in Fig. 4(a), the two crossing bands near $E_{F}$ mainly consist of set $\{d_{xy},d_{x^{2}-y^{2}}\}$. The degeneracy of this set is lifted by SOC. According to the first-order perturbation, the splitting can be written as

\begin{equation}
  \Delta E = (2\lambda \hbar^{2}+\frac{U_{eff}}{2\hbar}\vert\bar{L}_{z}\vert)\vert\cos \theta\vert,
\end{equation}
where $\theta$ is the angle between the magnetic moment and z-axis. The equation above indicates that the band gap of TiTe is independent of in-plane rotation angle $\phi$ (the angles $\theta$ and $\phi$ are illustrated in Fig. S5) and is zero with in-plane magnetization to the first order perturbation. The U-enhancing band gap increases as $\theta$ decreases until it reaches the maximum at $\theta=0$, in good agreement with the calculation results shown in Fig. 5(d).

It is interesting to note that as U increases, the easy magnetization axis shifts from in-plane to z-axis, followed by a rapid increase in MAE, as shown in Fig. 5(c). DOSs in Fig. 4(c) show that the occupied states are mostly $d_{z^{2}}$ and $d_{xy}$ in spin-up channel and are separated from unoccupied states by large crystal field splitting, resulting in first-order perturbation having negligible effect on single-ion anisotropy, and therefore we considered second-order perturbation for correlation-enhanced MAE as

\begin{equation}
  \begin{aligned}
    MAE &=(\frac{\frac{3}{4}\lambda_{\perp}^{2}}{\Delta_{\perp}}- \frac{\lambda_{\parallel}^{2}}{\Delta_{\parallel}})\hbar^{4}\cos^{2}\theta,
    \end{aligned}
\end{equation}
where $\Delta_{\parallel}=\vert e_{x^{2}-y^{2}}-e_{xy}\vert$ and $\Delta_{\perp}=\vert e_{z^{2}}-e_{yz}\vert$ are energy splitting for the in-plane and out-of-plane orbitals, respectively, which can be estimated by the weight-center positions of the DOSs \cite{Magnetism_Nanoscale2006,SOC-U,HO-LU-samSpin}. One can see from the equation above that for small U, the orbital moment quenched and the MAE are mainly controlled by the denominator. Because of the $\Delta_{\parallel}>\Delta_{\perp}$ for the small U (see Fig. 4(c)), $\left(\frac{\frac{3}{4}\lambda_{\perp}^{2}}{\Delta_{\perp}}-\frac{\lambda_{\parallel}^{2}}{\Delta_{\parallel}}\right)>0$ leads to an in-plane easy magnetization. Increased correlation enhances orbital polarization $\Delta L = \vert L_z \vert - \vert L_x \vert$ (Fig. 5(c)), and also improves $\Delta_{\perp}$ to exceed $\Delta_{\parallel}$ (Fig. 4(c)), resulting in $\left(\frac{\frac{3}{4}\lambda_{\perp}^{2}}{\Delta_{\perp}}-\frac{\lambda_{\parallel}^{2}}{\Delta_{\parallel}}\right)<0$, hence, the easy magnetization shifting to the z-direction.

We have some remarks before closing. First, the proposed 2D HCWSs here are protected by in-plane screw and glide symmetry. Hence, the Weyl points are no longer protected by breaking these nonsymmorphic symmetries when the in-plane magnetization deviates from the in-plane axial. However, from equation (7), one can see that the energy gap opening is independent of in-plane rotating angle $\phi$ to the first-order perturbation. The calculation results show a tiny gap (\textasciitilde3.5 meV) when a in-plane rotating applied, which means that the half Weyl semimetal properties are well maintained for any in-plane magnetization. 

Second, the equation (8) indicates that the MAE of TiTe is also less relevant to in-plane rotating angle $\phi$, which is consistent with the calculation results shown in Fig. 3(c). This property ensures that the in-plane magnetization of TiTe, as well as the chirality of its Weyl points, are easily modulated by the magnetic field.

Third, the out-of-plane magnetization will open a finite energy gap as shown in Fig. 5(d) and equation (7), whose Chern number is related with the magnetization direction. Thus, the chirality of edge states is locked with their spin direction, enabling more diverse modulation of TiTe. 

In summary, we have proposed a special topological state: 2D HCWSs. It represents a half Weyl semimetal bound to a half Chern insulator by symmetry. We show the symmetry conditions to realize this state in 2D systems with multiple screw axes. Futhermore, we show examples for 2D HCWSs based on TB models and a new material TiTe, a 2D FM magnet with a high $T_C$. Correlation-enhanced SOC and its inherent relevance to the MAE and energy gap opening in the TiTe are shown. These results demonstrate a universal proposal for realizing new low-power topological spintronics devices with easy manipulation and diverse tunable tools.

\textit{Computational methods }---Our first-principles density functional theory (DFT) calculations are carried out with the generalized gradient approximation proposed by Perdew, Burke, and Ernzerhof (PBE) \cite{PBE_GGA}, which is implemented in the Vienna ab initio simulation package (VASP) \cite{VASP}. The GGA+U method \cite{GGA+U} is employed to describe the strongly correlated Ti-$3d$ electrons, unless explicitly stated otherwise, all of the calculations were done with U=3 eV \cite{Ti_U3_1,Ti_U3_2}. The plane-wave cutoff energy was set to 500 eV and the vacuum space is more than 15 $\mathring{\text{A}}$ to avoid the influence between two adjacent slabs. The force converge criterion was less than 0.01 eV/$\mathring{\text{A}}$, the energies were less than $10^{-6}$ eV, and Monkhorst-Pack k-point grids of $12\times 12\times 1$ were adopted. The phonon spectra is calculated by using density functional perturbation
theory (DFPT) implemented in PHONOPY code \cite{phonopy_1,phonopy_2} with $3\times 3\times 1$ supercell. For the monolayer TiTe, the hybrid
functional HSE06 \cite{HSE06} is employed to verify band structure. The tight-binding model and the topology characteristics of TiTe monolayer are calculated by wannier90 \cite{wannier90} and wanniertools \cite{wanniertools}. To estimate the $T_C$ of the FM states, the Monte Carlo simulations \cite{Monte_Carlo_Tc} are performed on a $51\times 51$ supercell with $10^6$ steps at each temperature.

\begin{acknowledgments}
  This work was supported by National Natural Science Foundation of China under Nos. 11904101, 11604134, 11874117, 12174059 and the Natural Science Foundation of Shanghai under Grant No. 21ZR1408200.
\end{acknowledgments}
\bibliographystyle{unsrtnat}
\bibliography{TiTe}

\begin{thebibliography}{45}
\providecommand{\natexlab}[1]{#1}
\providecommand{\url}[1]{\texttt{#1}}
\expandafter\ifx\csname urlstyle\endcsname\relax
  \providecommand{\doi}[1]{doi: #1}\else
  \providecommand{\doi}{doi: \begingroup \urlstyle{rm}\Url}\fi

\bibitem[Zhou et~al.(2019)Zhou, Zhang, Zhang, Ma, Feng, Mokrousov, and
  Yao]{SpinPolarized_Nodalloop2019}
Xiaodong Zhou, Run-Wu Zhang, Zeying Zhang, Da-Shuai Ma, Wanxiang Feng, Yuriy
  Mokrousov, and Yugui Yao.
\newblock \emph{J. Phys. Chem. Lett.}, 10:\penalty0 3101--3108, 2019.

\bibitem[You et~al.(2019)You, Chen, Zhang, Sheng, Yang, and
  Su]{2D_WeylHalf_shengyuan_yang}
Jing-Yang You, Cong Chen, Zhen Zhang, Xian-Lei Sheng, Shengyuan~A. Yang, and
  Gang Su.
\newblock \emph{Phys. Rev. B}, 100:\penalty0 064408, 2019.

\bibitem[Zhang et~al.(2021)Zhang, Zhou, Zhang, Ma, Yu, Feng, and
  Yao]{WeylMonoloop_Half}
Run-Wu Zhang, Xiaodong Zhou, Zeying Zhang, Da-Shuai Ma, Zhi-Ming Yu, Wanxiang
  Feng, and Yugui Yao.
\newblock \emph{Nano Lett.}, 21:\penalty0 8749--8755, 2021.

\bibitem[Andrews et~al.(2017)Andrews, Fan, Forward, Chen, and
  Loock]{2D_WeylHalf_NiCS3}
N.~L.~P. Andrews, J.~Z. Fan, R.~L. Forward, M.~C. Chen, and H.-P. Loock.
\newblock \emph{Phys. Chem. Chem. Phys.}, 19:\penalty0 73--81, 2017.

\bibitem[Xue et~al.(2018)Xue, Zhao, Zhu, Zhou, Zhang, Li, Jiang, and
  Yang]{NPG_halfChernInsulators2018}
Yang Xue, Bao Zhao, Yan Zhu, Tong Zhou, Jiayong Zhang, Ningbo Li, Hua Jiang,
  and Zhongqin Yang.
\newblock \emph{NPG Asia Mater}, 10:\penalty0 e467--e467, 2018.

\bibitem[Huan et~al.(2021)Huan, Xue, Zhao, Gao, Bao, and
  Yang]{Half_QAH_Valley_Huanhao}
Hao Huan, Yang Xue, Bao Zhao, Guanyi Gao, Hairui Bao, and Zhongqin Yang.
\newblock \emph{Phys. Rev. B}, 104:\penalty0 165427, 2021.

\bibitem[Sun et~al.(2020)Sun, Ma, and Kioussis]{FeI_MAE_strain}
Qilong Sun, Yandong Ma, and Nicholas Kioussis.
\newblock \emph{Mater. Horiz.}, 7:\penalty0 2071--2077, 2020.

\bibitem[Li et~al.(2020)Li, Li, Li, Ye, Zheng, Zhang, Fu, Duan, and
  Xu]{xuyong_QAH_LiFeSe_2020}
Yang Li, Jiaheng Li, Yang Li, Meng Ye, Fawei Zheng, Zetao Zhang, Jingheng Fu,
  Wenhui Duan, and Yong Xu.
\newblock \emph{Phys. Rev. Lett.}, 125:\penalty0 086401, 2020.

\bibitem[{Lei Liu, Hao Huan, Yang Xue, Hairui Bao, {and} Zhongqin
  Yang}(2022)]{Ni2I2_Liulei}
{Lei Liu, Hao Huan, Yang Xue, Hairui Bao, {and} Zhongqin Yang}.
\newblock \emph{Nanoscale, to be published}, 2022.

\bibitem[Mermin and Wagner(1966)]{Mermin_Wagner_theorem}
N.~D. Mermin and H.~Wagner.
\newblock \emph{Phys. Rev. Lett.}, 17:\penalty0 1133--1136, 1966.

\bibitem[Li et~al.(2022)Li, Yao, Wu, Hu, Gao, Wan, and Liu]{U_enhanced_SOC_MAE}
Jiayu Li, Qiushi Yao, Lin Wu, Zongxiang Hu, Boya Gao, Xiangang Wan, and Qihang
  Liu.
\newblock \emph{Nat Commun}, 13:\penalty0 919, 2022.

\bibitem[Fan et~al.(2018)Fan, Ma, Fu, Liu, and Yao]{Cat_cradle_Dirac}
Xiaotong Fan, Dashuai Ma, Botao Fu, Cheng-Cheng Liu, and Yugui Yao.
\newblock \emph{Phys. Rev. B}, 98:\penalty0 195437, 2018.

\bibitem[Hirschmann et~al.(2021)Hirschmann, Leonhardt, Kilic, Fabini, and
  Schnyder]{abc}
Moritz~M. Hirschmann, Andreas Leonhardt, Berkay Kilic, Douglas~H. Fabini, and
  Andreas~P. Schnyder.
\newblock \emph{Phys. Rev. Materials}, 5:\penalty0 054202, 2021.

\bibitem[Furusaki(2017)]{Weyl_Dirac_Screw}
Akira Furusaki.
\newblock \emph{Science Bulletin}, 62:\penalty0 788--794, 2017.

\bibitem[SMs(2022)]{SMs}
\emph{Supplemental Materials}, 2022.

\bibitem[Nielsen and Ninomiya(1983)]{no_go_theorem}
H.~B. Nielsen and Masao Ninomiya.
\newblock \emph{Physics Letters B}, 130:\penalty0 389--396, 1983.

\bibitem[Yao et~al.(2009)Yao, Yang, and Niu]{half_topo_charge_Weyl_point}
Wang Yao, Shengyuan~A. Yang, and Qian Niu.
\newblock \emph{Phys. Rev. Lett.}, 102:\penalty0 096801, 2009.

\bibitem[Wan et~al.(2011)Wan, Turner, Vishwanath, and
  Savrasov]{wanxiangang_Weyl_SemimetalFermiarc2011}
Xiangang Wan, Ari~M. Turner, Ashvin Vishwanath, and Sergey~Y. Savrasov.
\newblock \emph{Phys. Rev. B}, 83:\penalty0 205101, 2011.

\bibitem[Zhang et~al.(2022)Zhang, Yu, Liu, and Yao]{MagneticTB}
Zeying Zhang, Zhi-Ming Yu, Gui-Bin Liu, and Yugui Yao.
\newblock \emph{Computer Physics Communications}, 270:\penalty0 108153, 2022.

\bibitem[Yu and Trinkle(2011)]{Bader_charge}
Min Yu and Dallas~R. Trinkle.
\newblock \emph{The Journal of chemical physics}, 134:\penalty0 064111, 2011.

\bibitem[Goodenough(1976)]{goodenough_kanamori_anderson}
J.B. Goodenough.
\newblock Interscience Monographs on Chemistry, Inorganic Chemistry Section.
  {Krieger}, 1976.
\newblock https://books.google.com/books?id=2sjdtAEACAAJ.

\bibitem[Anderson(1959)]{GKA_A}
P.~W. Anderson.
\newblock \emph{Phys. Rev.}, 115:\penalty0 2--13, 1959.

\bibitem[Kanamori(1960)]{GKA_K}
Junjiro Kanamori.
\newblock \emph{Journal of Applied Physics}, 31:\penalty0 S14--S23, 1960.

\bibitem[Goodenough(1955)]{GKA_G}
John~B. Goodenough.
\newblock \emph{Phys. Rev.}, 100:\penalty0 564--573, 1955.

\bibitem[St{\"o}hr and Siegmann(2006)]{Magnetism_Nanoscale2006}
Joachim St{\"o}hr and Hans~Christoph Siegmann.
\newblock Springer Series in Solid-State Sciences. {Springer}, {Berlin ; New
  York}, 2006.
\newblock ISBN 978-3-540-30282-7.

\bibitem[Cheng et~al.(2021)Cheng, Xu, Jia, Zhao, Hu, Wu, and Ren]{VP_weiren}
Xuli Cheng, Shaowen Xu, Fanhao Jia, Guodong Zhao, Minglang Hu, Wei Wu, and Wei
  Ren.
\newblock \emph{Phys. Rev. B}, 104:\penalty0 104417, 2021.

\bibitem[Huang et~al.(2017)Huang, Clark, {Navarro-Moratalla}, Klein, Cheng,
  Seyler, Zhong, Schmidgall, McGuire, Cobden, Yao, Xiao, {Jarillo-Herrero}, and
  Xu]{CrI3_45K}
Bevin Huang, Genevieve Clark, Efr{\'e}n {Navarro-Moratalla}, Dahlia~R. Klein,
  Ran Cheng, Kyle~L. Seyler, Ding Zhong, Emma Schmidgall, Michael~A. McGuire,
  David~H. Cobden, Wang Yao, Di~Xiao, Pablo {Jarillo-Herrero}, and Xiaodong Xu.
\newblock \emph{Nature}, 546:\penalty0 270--273, 2017.

\bibitem[Zhang et~al.(2015)Zhang, Qu, Zhu, and Lam]{CrI3_95K}
Wei-Bing Zhang, Qian Qu, Peng Zhu, and Chi-Hang Lam.
\newblock \emph{J. Mater. Chem. C}, 3:\penalty0 12457--12468, 2015.

\bibitem[Yang et~al.(2015)Yang, Liu, Sun, Peng, Yang, Zhang, Zhou, Zhang, Guo,
  Rahn, Prabhakaran, Hussain, Mo, Felser, Yan, and Chen]{exp_TaAs_Weylsemi}
L.~X. Yang, Z.~K. Liu, Y.~Sun, H.~Peng, H.~F. Yang, T.~Zhang, B.~Zhou,
  Y.~Zhang, Y.~F. Guo, M.~Rahn, D.~Prabhakaran, Z.~Hussain, S.-K. Mo,
  C.~Felser, B.~Yan, and Y.~L. Chen.
\newblock \emph{Nature Phys}, 11:\penalty0 728--732, 2015.

\bibitem[Weng et~al.(2015)Weng, Fang, Fang, Bernevig, and
  Dai]{TaAs_WeylSemi_Weng}
Hongming Weng, Chen Fang, Zhong Fang, B.~Andrei Bernevig, and Xi~Dai.
\newblock \emph{Phys. Rev. X}, 5:\penalty0 011029, 2015.

\bibitem[Xu et~al.(2011)Xu, Weng, Wang, Dai, and Fang]{HgCr2Se4_WyelSemi}
Gang Xu, Hongming Weng, Zhijun Wang, Xi~Dai, and Zhong Fang.
\newblock \emph{Phys. Rev. Lett.}, 107:\penalty0 186806, 2011.

\bibitem[Liu et~al.(2022)Liu, Zhang, Yu, Yang, and Yao]{Magnetic_III}
Gui-Bin Liu, Zeying Zhang, Zhi-Ming Yu, Shengyuan~A. Yang, and Yugui Yao.
\newblock \emph{Phys. Rev. B}, 105:\penalty0 085117, 2022.

\bibitem[Wang et~al.(2017)Wang, Tang, Du, and Wan]{SOC-U}
Di~Wang, Feng Tang, Yongping Du, and Xiangang Wan.
\newblock \emph{Phys. Rev. B}, 96:\penalty0 205159, 2017.

\bibitem[Whangbo et~al.(2015)Whangbo, Gordon, Xiang, Koo, and
  Lee]{HO-LU-samSpin}
Myung-Hwan Whangbo, Elijah~E. Gordon, Hongjun Xiang, Hyun-Joo Koo, and
  Changhoon Lee.
\newblock \emph{Acc. Chem. Res.}, 48:\penalty0 3080--3087, 2015.

\bibitem[Perdew et~al.(1996)Perdew, Burke, and Ernzerhof]{PBE_GGA}
John~P. Perdew, Kieron Burke, and Matthias Ernzerhof.
\newblock \emph{Phys. Rev. Lett.}, 77:\penalty0 3865--3868, 1996.

\bibitem[Kresse and Furthm{\"u}ller(1996)]{VASP}
G.~Kresse and J.~Furthm{\"u}ller.
\newblock \emph{Phys. Rev. B}, 54:\penalty0 11169--11186, 1996.

\bibitem[Liechtenstein et~al.(1995)Liechtenstein, Anisimov, and Zaanen]{GGA+U}
A.~I. Liechtenstein, V.~I. Anisimov, and J.~Zaanen.
\newblock \emph{Phys. Rev. B}, 52:\penalty0 R5467--R5470, 1995.

\bibitem[Lutfalla et~al.(2011)Lutfalla, Shapovalov, and Bell]{Ti_U3_1}
Suzanne Lutfalla, Vladimir Shapovalov, and Alexis~T. Bell.
\newblock \emph{J. Chem. Theory Comput.}, 7:\penalty0 2218--2223, 2011.

\bibitem[Hu and Metiu(2011)]{Ti_U3_2}
Zhenpeng Hu and Horia Metiu.
\newblock \emph{J. Phys. Chem. C}, 115:\penalty0 5841--5845, 2011.

\bibitem[Togo et~al.(2008)Togo, Oba, and Tanaka]{phonopy_1}
Atsushi Togo, Fumiyasu Oba, and Isao Tanaka.
\newblock \emph{Phys. Rev. B}, 78:\penalty0 134106, 2008.

\bibitem[Togo and Tanaka(2015)]{phonopy_2}
Atsushi Togo and Isao Tanaka.
\newblock \emph{Scripta Materialia}, 108:\penalty0 1--5, 2015.

\bibitem[Heyd et~al.(2003)Heyd, Scuseria, and Ernzerhof]{HSE06}
Jochen Heyd, Gustavo~E. Scuseria, and Matthias Ernzerhof.
\newblock \emph{J. Chem. Phys.}, 118:\penalty0 8207--8215, 2003.

\bibitem[Mostofi et~al.(2008)Mostofi, Yates, Lee, Souza, Vanderbilt, and
  Marzari]{wannier90}
Arash~A. Mostofi, Jonathan~R. Yates, Young-Su Lee, Ivo Souza, David Vanderbilt,
  and Nicola Marzari.
\newblock \emph{Computer Physics Communications}, 178:\penalty0 685--699, 2008.

\bibitem[Wu et~al.(2018)Wu, Zhang, Song, Troyer, and Soluyanov]{wanniertools}
QuanSheng Wu, ShengNan Zhang, Hai-Feng Song, Matthias Troyer, and Alexey~A.
  Soluyanov.
\newblock \emph{Computer Physics Communications}, 224:\penalty0 405--416, 2018.

\bibitem[Evans et~al.(2014)Evans, Fan, Chureemart, Ostler, Ellis, and
  Chantrell]{Monte_Carlo_Tc}
R.~F.~L. Evans, W.~J. Fan, P.~Chureemart, T.~A. Ostler, M.~O.~A. Ellis, and
  R.~W. Chantrell.
\newblock \emph{J. Phys.: Condens. Matter}, 26:\penalty0 103202, 2014.

\end{thebibliography}
\end{document}